\documentclass[mlabstract,twocolumn]{jmlr}

\usepackage{longtable}

\usepackage{booktabs}
\usepackage[load-configurations=version-1]{siunitx} 

\theorembodyfont{\upshape}
\theoremheaderfont{\scshape}
\theorempostheader{:}
\theoremsep{\newline}

\title[  
ML Case Study for AI-empowered echocardiography of ICU Patients in LMIC 
]{
A Machine Learning Case Study for AI-empowered echocardiography of Intensive Care Unit Patients in low- and middle-income countries  
}

  \author{\Name{Miguel Xochicale} \Email{m.xochicale@ucl.ac.uk}\\
  \addr ARC/WEISS University College London
  \AND
  \Name{Louise Thwaites} \Email{lthwaites@oucru.org}\\
  \Name{Sophie Yacoub}  \Email{syacoub@oucru.org}\\
  \addr Oxford University Clinical Research Unit Vietnam
  \AND
  \Name{Luigi Pisani}  \Email{luigipisani@gmail.com}\\
  \addr Mahidol Oxford Tropical Medicine Research Unit
  \AND
  \Name{Nhat Phung Tran Huy}  \Email{nhat.phung@kcl.ac.uk}\\
  \Name{Hamideh Kerdegari}  \Email{hamideh.kerdegari@kcl.ac.uk}\\
  \Name{Andrew King} \Email{andrew.king@kcl.ac.uk}\\
  \Name{Alberto Gomez} \Email{alberto.gomez@kcl.ac.uk}\\
  \addr BMEIS, King's College London
 }

\begin{document}

\maketitle

\begin{abstract}
We present a Machine Learning (ML) study case to illustrate the challenges of clinical translation for a real-time AI-empowered echocardiography system with data of ICU patients in LMICs.
Such ML case study includes data preparation, curation and labelling from 2D Ultrasound videos of 31 ICU patients in LMICs and model selection, validation and deployment of three thinner neural networks to classify apical four-chamber view.
Results of the ML heuristics showed the promising implementation, validation and application of thinner networks to classify 4CV with limited datasets.
We conclude this work mentioning the need for (a) datasets to improve diversity of demographics, diseases, and (b) the need of further investigations of thinner models to be run and implemented in low-cost hardware to be clinically translated in the ICU in LMICs.
\end{abstract}

\section{Introduction}
\label{sec:intro}
Echocardiography is an important clinical procedure in Intensive Care Units (ICUs) because of the features of Ultrasound (US) image modality such as portability, low cost, non-ionising radiation and its real-time capabilities to visualise cardiac anatomy~\citep{Vieillard-Baron2008, cambell2018}. 
Typically, the identification of cardiac abnormalities from 2D US views (Apical 4-Chamber (A4C), Parastemal Long-Axis and other six views) is achieved by specialist clinicians in echocardigraphy following the Focused Intensive Care Echo protocol~\citep{2017_hall_JIntensiveCareSociety}. 
However, the application of point-of-care echocardiography of ICU patients in low- and middle-income countries (LMICs) faces two challenges:
(1) intra-view variability of echocardiograms (physiological variations of patients and acquisition parameters) and inter-observer variability of expertise for sonographer and radiologist~\citep{Feigenbaum1996, khamis2017, field2011}, and
(2) limited number of specialist clinicians to perform US imaging analysis and to provide accurate diagnosis, and the limited equipment and hospitalisation requirements~\citep{2016_becker_in_TropicalMedicineInternationalHealth, hao2021-wellcome, 2021-huyNhat-vanHao-in-FAIR-MICCAI}.
One promising approach to address such challenges is the application of Machine Learing (ML) and Artificial Intelligence (AI) methods to echocardiography~\citep{2022asch_AmericanSocietyofEchocardiography}.
AI-empowered echocardiography has been successful for detection of different apical views, inter-observer variability of sonographer's expertise, implementation of one-stop AI models with multimodal imaging (US, MRI and clinical data), detection of high/low risk of heart failure, detection of endocardial borders and automatic left ventricle assessment in 2D echocardiography videos~\citep{tromp2022, zhang2022-mdpi, behnami2020, ono2022}.
In spite of the success in applying AI and ML methods to support echocardiography, there are still important challenges for these methods to be integrated as clinical system and translated to clinical practice:
\begin{enumerate}
\setlength\itemsep{0em}
\item inter-view similarity of echocardiograms (similar views of valve motion, wall motion, left ventricle, etc) and transducer position during acquisition when performing serial echoes~\citep{zhang2018},
\item redundant information in the clinical echo system (icons, date, frame rate, etc)~\citep{khamis2017} and variation of US images from different clinical US systems~\citep{brindise2020unsupervised}, and
\item internal and external validation of AI-based models, data patient privacy to train commercial algorithms, and regulations of software as medical devices~\citep{2022_Stewart_Emergency_Medicine_Australasia}.
\end{enumerate}
The first and second challenges are important because of 2D US video data requires to appropriately be collected, validated and managed to apply AL and ML methods, and the third challenge because AI-based medical devices require to be aligned to standards to then be ready for clinical translation.
Hence, the adoption of good machine learning practices (data curation, open-source code implementation, model selection, training and tuning; model validation and inference) might help to address such challenges.

This work presents a scoping review for (a) AI-empowered echocardiography for ICU in LMICs and (b) classification US images with thinner neural networks.
We present a ML case study to classify A4C US image considering three thinner neural networks trained with curated data from ICU patients in LMICs.
We present training results from SqueezeNet for different datasize, number of frames and clips batches. 
Further details for datsets, results of thinner networks are also added in the appendixes.
We conclude with needs for clinical translation and add future work with validation of AI-based models. 

\section{Scoping review}
\subsection{AI-empowered echocardiography for ICU in LMICs}
\citet{hanson2001} reviewed various AI-based applications in the ICU where real-time analysis of waveforms of electrocardiograms and electroencephalograms with neural network were used to identify cardiac ischemia and diagnosis of myocardial ischemia.
\citet{Ghorbani-DigitalMedicineNature-JAN2020} reported how deep learning models predicts systematic phenotypes from echocardiogram images which are difficult for human interpreters.
\citet{CHEEMA2021JACCCaseReports} reported five patients with covid-19 in the ICU to illustrate "how decision making affect in patient care" and how the use of AI-enabled tools provided real-time guidance to acquire desired cardiac 2D US views.
Recently, \citet{hong2022} reviewed 673 papers that apply ML methods to help making clinical decision in the ICU, of these studies the majority used supervised learning (91\%) and few of them applied unsupervised learning and reinforcement learning methods.
Similarly, \citet{hong2022} identified 20 of the most frequent variables in ML pipelines for ICU patients, being the top five (age, sex, heart rate, respiratory rate, and pH), 
and mentionted that the most studied diseases are sepsis, infection and kidney injury.
Despite such advances, there is little research on AI-empowered echocardiography used by clinicians in the ICU, specifically in LMICs.
For instance, \cite{2021-huyNhat-vanHao-in-FAIR-MICCAI} reported challenges in resourced limited ICUs including: infrastructure, education, personnel, data pipelines, regulation and trust in AI.
\cite{2021-kerdegari-Applied-Sciences-MDPI, 2021-kerdegari-ISBI-IEEE, 2021-huyNhat-kerdegari-in-FAIR-MICCAI} presented a deep-learning pipelines to classify lung US pathologies for ICU patients in LMICs, stating the challenges of data imbalance, integration of technology and the limited IT infrastructure.

\subsection{Classification of US images with thinner neural networks} \label{subsec:thinnerNets}
Classifying echochardiograms has been sucessfully done in the previous years (\sectionref{subsec:nns_echochardiograms}), however real-time deployment of AI-models suggest to require thinner networks.
For instance, \citet{baumgartner2017-IEEETransMedImag} proposed SonoNet which is a VGG-based architecture, having the same first 13 layers of VGG16, and SmallNet, loosely inspired by AlexNet, for real-time detection and bounding box localisation of standard views in freehand fetal US.
\citet{toussaint2018-MICCAI} applied four feature extraction networks couple with batchnormalization and soft proposal layer (VGG13-SP, VGG16-SP, ResNet18-SP, ResNet34-SP), resulting in real-time performance at inference time of 40 m$s$ per image ($\sim$20Hz)
\citet{Al-Dhabyani2019-IJACSA} applied AlexNet and transfer learning of four architectures (VGG16, Inception, ResNet, and NASNet) with and without augmentation techniques to perform tumor classification of breast US imaging.
Authors stated that transfer learning with NASNet presented the best accuracy (99\%) using BUSI+B datasets with DAGAN augmentation.
\citet{xie2020-physics-in-medicine-biology} proposed a dual-sampling convolutional neural network for US image breast cancer classification, being their network more efficient for such task than AlexNet, VGG16, ResNet18, GoogleNet and EfficientNet.
Recently, \citet{snider2022-ScientificReports} reported summaries of CNN heuristics to detect shrapnel in US images, including layer activators, 2D CNN layer architectures, model optimisers dense nodes, and the effect of image augmentation and dropout rate and epoch number.
Similarly, \citet{boice2022-in-jimaging} proposed ShrapML, a CNN model to detect shrapnel in US imaging.
Authors compared ShrapML against DarkNet19, GoogleNet, MobileNetV2 and SqueezeNet, being ShrapML (8layers--6CNN and 2FC, making a network of 0.43 million parameters) 10x faster than MobileNetV2 and offering the highest accuracy.

\section{Machine learning case study}
\begin{figure}[ht]
\floatconts
  {fig:main-figure}
  {\caption{
      ML pipeline of real-time AI-empowered echocardiography for ICU in LMICs:
      (a) timestamp labelling of A4C frames and clips,
      (b) classification pipeline with thinner neural networks, and
      (c) low-cost clinical system with Epiq Q7, cardiac probe X5-1, USB video-frame grabber and 16GB GeForce RTX 3080 GPU Laptop.
    }
  }
    {\includegraphics[width=0.70\linewidth]{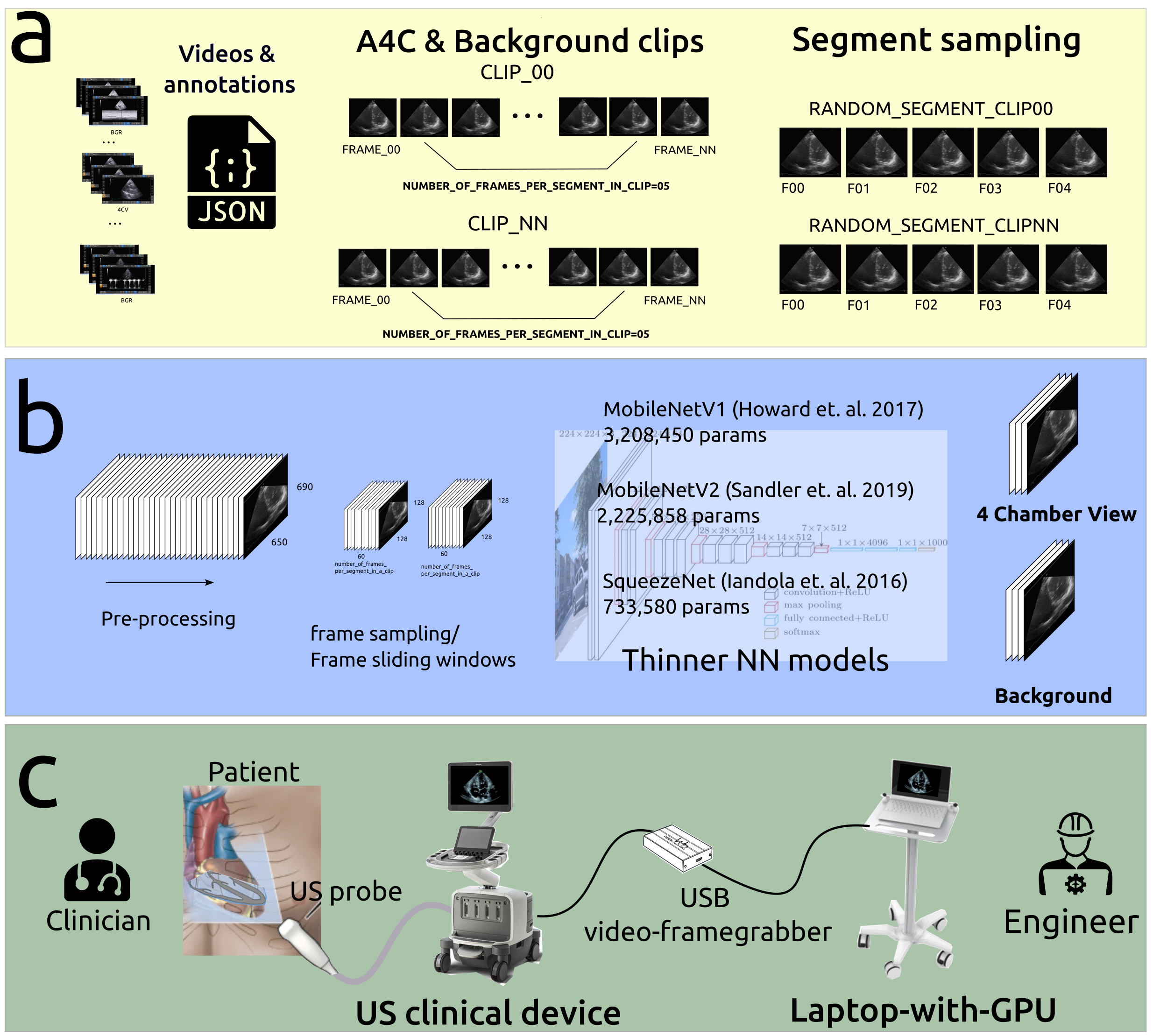}} 
\end{figure}
\subsection{Dataset}
Echocardiography videos of 31 anonymised patients in the ICU were considered for this work which were collected by four radiologists using the clinical US devices: GE Venue Go machine with GE convex probe C1-5-D.
The 31 patients had the following demographics:
Sex: \% (Male): 58.1\%;
Age: 38.70 (16.08) years;
Weight: 61.51 (15.06) Kg;
Height: 1.62 (0.07) m, and 
BMI: 23.80 (4.30). 
Values represent mean(std).
See \appendixref{apd:datasets} for further details on the demographics of the dataset, including the complete dataset of 87 patients.

\subsubsection{Ethics statement}
This study was approved by the anonymised entity one and anonymised entity two.
All participants gave written informed consent to participate in the data collection before enrollment.

\subsubsection{Data annotation, validation and management}
Apical 4 Chamber view (A4C) is considered as an important view to compute heart failure measurements from 2D US echocardiography ~\citep{2017_hall_JIntensiveCareSociety}.
Hence, for this work, timestamps in the video files for A4C were annotated by one research clinician of 10 years of experience using VGG Image Annotator (VIA).
Then the same clinician  and one researcher validated timestamps annotations where few filenames and timestamps were fixed (\figureref{fig:main-figure}(a)).

\subsection{Model selection and heuristics}
Considering different datasets characteristic (number of frames, clips, pixel image, clinical equipment, etc) and the number of parameters of different networks,
we selected three neural networks for this work:
MobileNetV1~\citep{2017-howared_CoRR_MobileNetV1} with 3,208,450 parameters, MobileNetV2~\citep{Sandler_2018_CVPR_MobileNetV2} with 2,225,858 parameters, and SqueezeNet~\citep{iandola2017squeezenet} with 733,580 parameters.
We then performed heuristics for each model to understand the impact of their performance for different hyperparameters (dataset size, frames numbers and clip length). 
MobileNetV1, MobileNetV2 and SqueezeNet were trained with data of 5 and 31 subjects (\appendixref{apd:heuristics}) in which SqueezeNet showed constant training metrics.
Hence, \figureref{fig:heuristics_results_SqueezeNet_source0} presents 
the impact of different datasize, batch size of clips, and number of frames per segment in a clip for SqueezeNet.
\begin{figure}[ht]
\floatconts
  {fig:heuristics_results_SqueezeNet_source0}
  {
      \caption{
          Heuristics for SqueezeNet with dataset of 5 subjects: 
          (a) varying batch size with constant number of frames per segment equal to 1, and
          (b) varying number of frames per clip with constant batch size of clips equal to 10.
      }
  }
    {\includegraphics[width=0.81\linewidth]{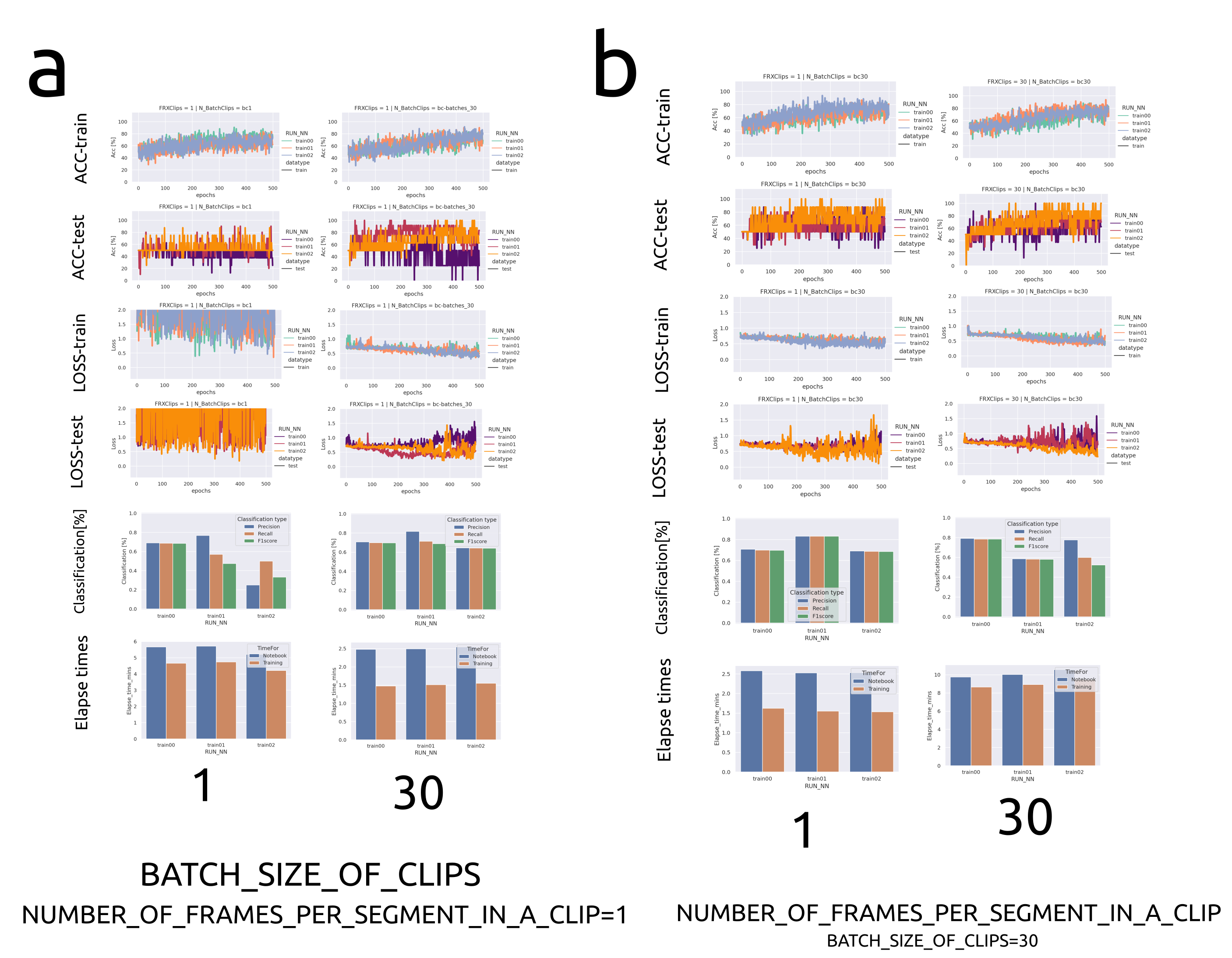}} 
\end{figure}

\section{Conclusions and Future Work}
We presented a ML case study that includes data selection, validation and management, model selection and validation.
Our results show the promising use of thinner neural networks to address scarcity of datasets and model deployment in low-cost hardware.
Future work will include real-time inference and deployment of thinner classification and segmentation models, leading to clinical translation of cardiac outputs of ICU patiens in LMICs.
The code, data and other resources to reproduce this work will be available upon acceptance of this work.



\newpage


\newpage
\appendix

\newpage
\section{Classification of echochardiograms} \label{subsec:nns_echochardiograms}
\citet{khamis2017} considered 309 clinical echocardiograms of apical views which were visually classified and labelled by two experts into three classes: 103 A2C views, 103 A4C views and 103 ALX views to then applied spatio-temporal feature extraction (Cuboic Detector) and supervised learning dictionary (LC-KSVD) resulting in an overall recognition rate of 95\%.
\citet{woudenberg2018} applied DenseNet and LSTM to extract temporal information on sequences of 16000 echo cine frames to classify 14 heart views with an average accuracy of 92.35\%.
\citet{woudenberg2018} also presents timing diagrams to quantify frame arrival and real-time performance to operate at 30 frames per second, while providing feedback with a mean latency of 352.91 ± 38.27 m$s$ when measured from the middle of the ten-frame sequence.
\citet{zhang2018} performed view classification with 277 echocardiograms to create a 23-class models (including A4C no occlusions, A4C occluded LA, A4C occluded LV, etc) using 13-layer CNN with 5-fold cross-validation for accuracy assessment and resulting in 84\% for overall accuracy where challenges for partial obscured LVs for A2C, A3C and A4C.
Similarly, \citet{zhang2018} applied U-net to segment 5 views (A2C, A3C, A4C, PSAX, PLAX) and CNN model for 3 cardiac diseases with the use of A4C capturing most of the information for the diseases.

\newpage
\section{Datasets}\label{apd:datasets}
\figureref{fig:demographics} illustrates demographics for sex, age, BMI, sepsis and dengue for the complete dataset and the 31 subjects considered for this work.
\begin{figure}[htbp]
\floatconts
  {fig:demographics}
  {\caption{
      Patient demographics for sex, age, BMI, sepsis and dengue disease.
      The total number of patient is 87 of which data from 31 were curated, annotated and validated.}} 
    {\includegraphics[width=\columnwidth]{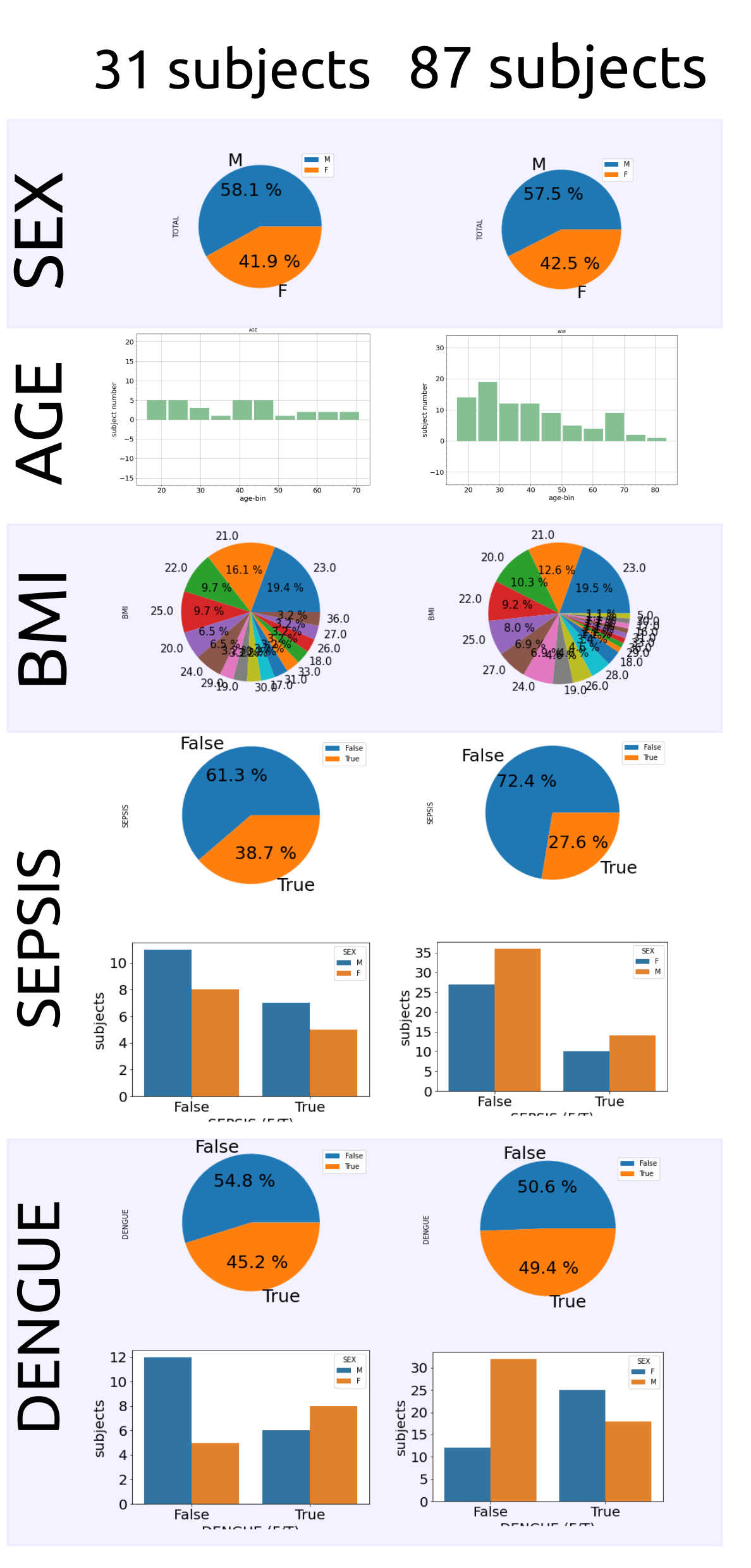}} 
\end{figure}

\section{Heuristics of model selection} \label{apd:heuristics}
\figureref{fig:ThinnerNeuralNetsResults} illustratres heristics for accuraty, train, classification and elapse times of 5 and 31 subjects.
\begin{figure*}[ht]
\floatconts
  {fig:ThinnerNeuralNetsResults}
  {\caption{
    Heuristics for 5 and 31 subjects with 1 frames per clip and 20 batch size of clips for
    MobileNetV1~\citep{2017-howared_CoRR_MobileNetV1} with 3,208,450 parameters, MobileNetV2~\citep{Sandler_2018_CVPR_MobileNetV2} with 2,225,858 parameters, and SqueezeNet~\citep{iandola2017squeezenet} with 733,580 parameters.
    Models were trained three timres (run01, run02 and run03) to illustratre the impact of the training on randomised datasets wich were splitted 70\% for train, 15\% for test and 15\% for validation.  
    \textbf{Note:} It is recommended to zoom in the figures to see further details (figure DPI is 300).
          }
  }
    {\includegraphics[width=\textwidth]{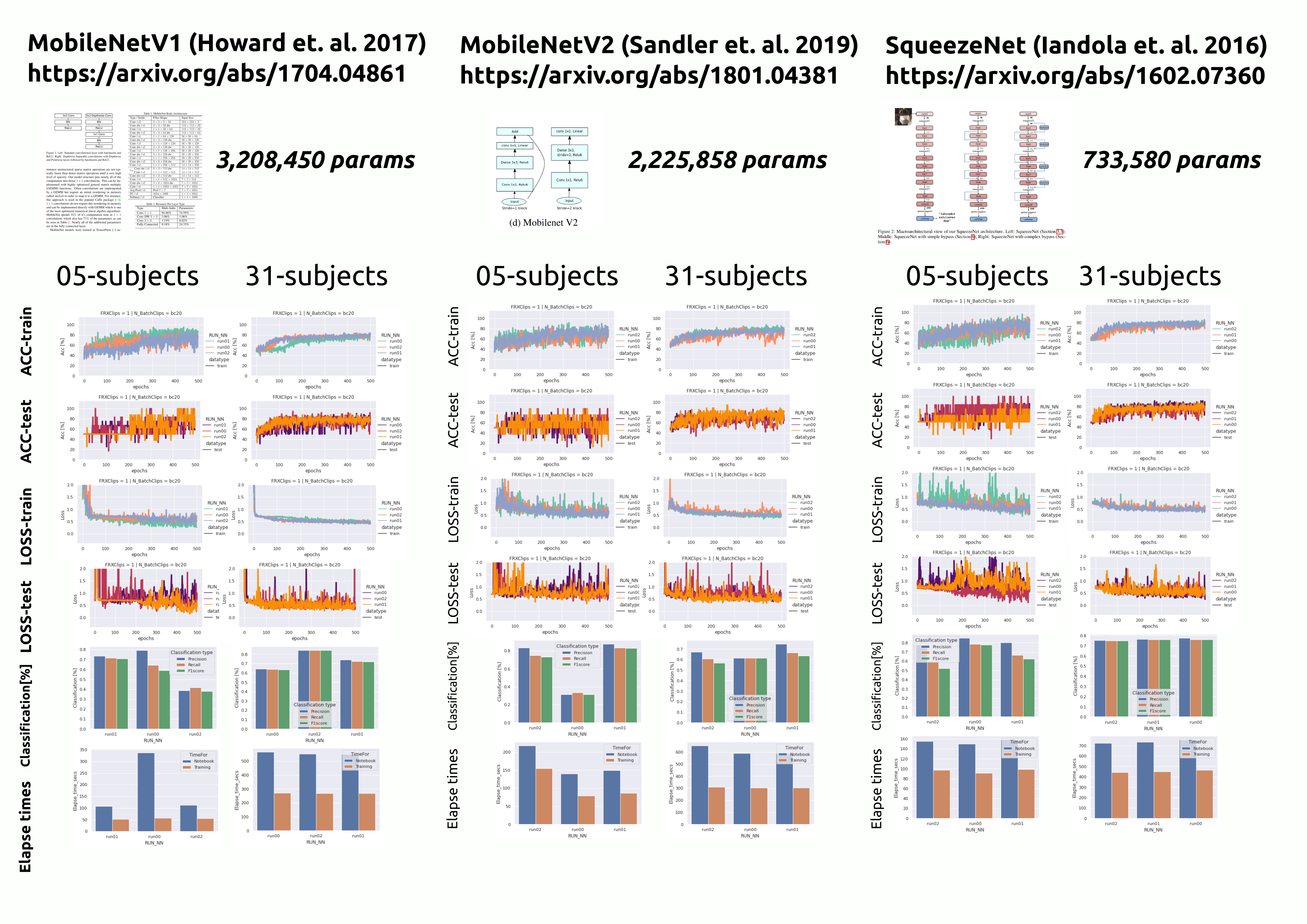}} 
\end{figure*}

\end{document}